\documentclass[aps,prb,twocolumn,eqsecnum,amsmath,amssymb,floatfix]{revtex4-1}
\usepackage{tabularx}
\usepackage{bm}
\usepackage{euscript}
\usepackage{epsfig,psfrag,subfigure}
\usepackage{graphicx}
\usepackage{color}
\usepackage{amsfonts}
\usepackage{exscale}
\usepackage{amsbsy}

\begin{document}

\title{Local Interlayer Tunneling Between Two-Dimensional Electron Systems in the Ballistic Regime}
\author{Katherine Luna, Eun-Ah Kim,\footnote{Permanent address Department of Physics, Cornell University, Ithaca, New York 14853, USA} Paul Oreto, Steven A. Kivelson, David Goldhaber-Gordon}
\affiliation{Department of Physics, Stanford University, Stanford CA 94305-4045, USA,} 
\affiliation{Stanford Institute for Materials and Energy Sciences}, 
\affiliation{SLAC National Accelerator Laboratory, 2575 Sand Hill Road, Menlo Park, CA 94025, USA}
\date{\today}

\begin{abstract}
We study a theoretical model of virtual scanning tunneling microscopy (VSTM)\cite{DGG,AdamTransistor}: a proposed application of interlayer tunneling in a bilayer system to locally probe a two-dimensional electron system (2DES) in a semiconductor heterostructure. We
consider tunneling for the case where transport in the 2DESs is ballistic,
and show that the zero-bias anomaly is suppressed by extremely efficient screening. Since such an anomaly would complicate the interpretation of data from VSTM, this result is encouraging for efforts to implement such a microscopy technique.
\end{abstract}

\maketitle

\section{Introduction}\label{Introduction}
The availability of increasingly clean low-density two-dimensional electronic systems (2DESs) has allowed access to a regime in which electron-electron interactions play a major role.\cite{AdamTransistor,DGG,Gao05,Hirjibehedin02} Evidence is accumulating from transport measurements that the physics of this regime is much richer than was previously appreciated (see Ref. \onlinecite{Abrahams} and references therein). In particular, while much is understood about the two limiting cases, $r_s \equiv 1/\sqrt{\pi n a_B^2}\rightarrow 0$ (Fermi liquid) and $r_s\rightarrow\infty$ (Wigner crystal), experiments on systems with intermediate values of $r_s = 10 - 30$ reveal a host of unanticipated anomalies.\cite{Abrahams,Spivak06} (Here $n$ is the
density of doped electrons or holes and $a_B$ is the effective Bohr radius, $a_B = \hbar^2\epsilon /m^*e^2$, where $m^*$ is the effective mass, $e$ is the electron charge, and $\epsilon$ is the dielectric constant of the host semiconductor.)

Experimental attempts to understand electron organization in 2DESs have been based mainly on transport measurements on large (micron to millimeter) scales. Direct information on the local structure of electronic states could powerfully elucidate the physics underlying these transport measurements, including the recently proposed ``electronic microemulsion phases.''\cite{Spivak04,Spivak06} Momentum-space probes\cite{Hirjibehedin02} and finite-frequency probes\cite{Sambandamurthy08} have provided
important insights, but the residual spatial inhomogeneity in even the cleanest low-density 2DESs favors the use of real-space probes. Over the past decade, important progress has been made in locally probing 2DESs.\cite{Yacoby,Topinka,Finkelstein, Ilani04, Woodside01, Jura07} However, a 2DES is generally buried $\sim$100 nanometers deep in a heterostructure, preventing the use of
scanning tunneling microscopy (STM), which has offered powerful insights into 2DESs at surfaces by mapping the local density of states at low energies.

\begin{figure}[h,t]
\includegraphics[width=0.4\textwidth]{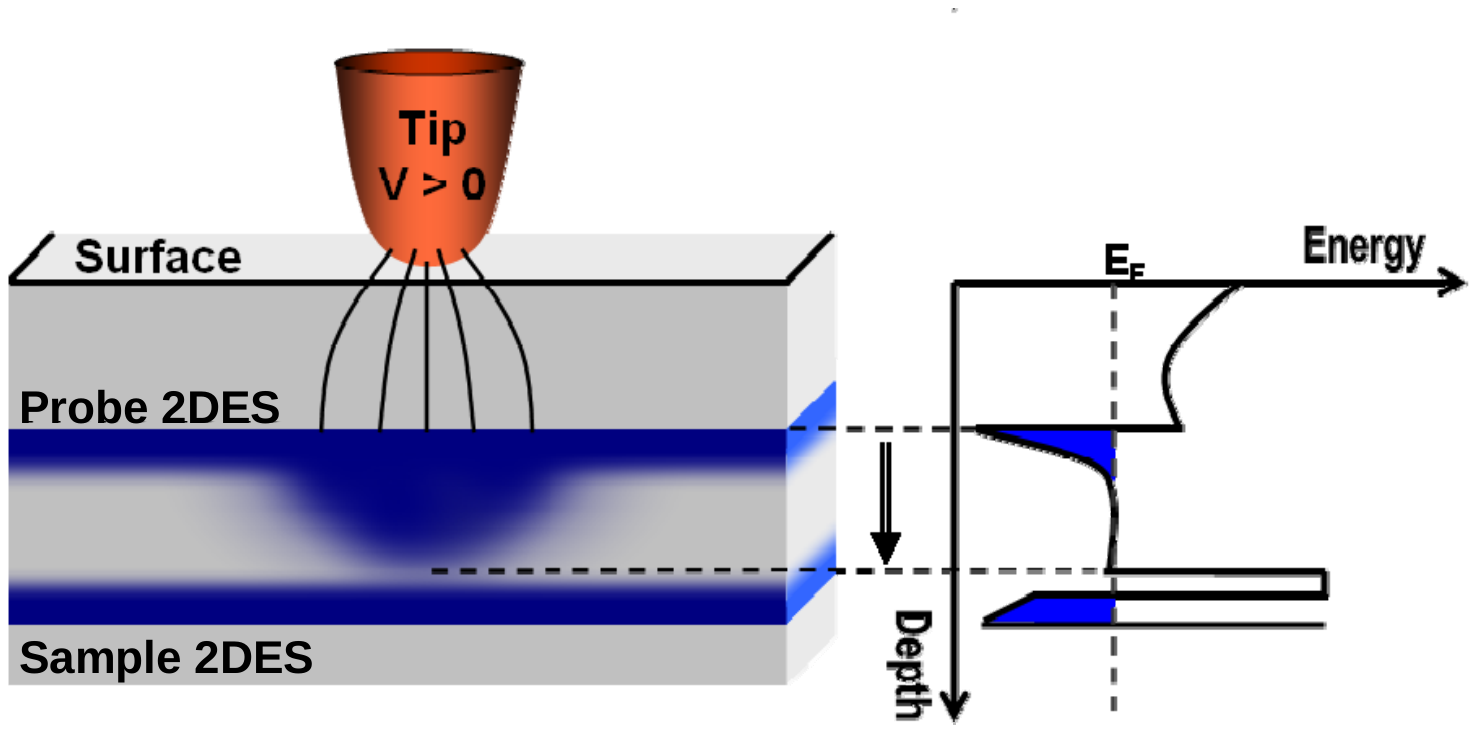} 
\caption{A schematic depiction of the proposed VSTM setup where an applied voltage by an actual tip induces a virtual tip in the probe 2DES which can then measure the sample 2DES via a tunneling process.  Figure courtesy of Adam Sciambi.\cite{DGG}}
\label{TunnExpt}
\end{figure}

An ongoing effort to develop a comparable technique for buried structures -- termed virtual scanning tunneling microscopy or VSTM\cite{DGG, AdamTransistor} -- is based on tunneling into a 2DES not from a scanned metal tip as in STM, but rather from a second ``Probe'' 2DES grown above the 2DES of interest (henceforth ``Subject 2DES''), within the same heterostructure. Since the barrier between the two 2DESs can be made very low by proper design of the layer structure, and since the Probe 2DES is not perfectly compressible, it should be possible to tune the barrier at a particular location by applying a voltage to a sharp metal tip positioned above the heterostructure surface (See Fig.\ref{TunnExpt}). Separate contact can be made to the Probe and Subject 
layers.\cite{Eisenstein90} Tunneling between the Probe and Subject 2DESs would then be strongly enhanced locally below the tip, and the location of enhanced tunneling could be scanned across the Sample 2DES by scanning the metal tip above the heterostructure. Such enhanced tunneling by over two orders of magnitude has recently been demonstrated by one of the present authors, though up to now the enhancement is over large areas as the tuning is done using a large-area gate rather than a sharp tip.\cite{DGG, AdamTransistor} Moving to a scannable tip will reduce signal size and require lockin detection techniques to separate local tunneling current from large area background tunneling. Work is underway to implement this.

In this paper, we introduce a minimal model for VSTM -- two parallel 2DESs connected by tunneling at a single point -- and use it to address the feasibility of VSTM at its simplest level. 
A VSTM setup should meet the following criteria, which our model must address: (i) there should be sufficient tunneling near zero bias to probe the low energy physics of interest, (ii) the tunneling rate should be sensitive to the {\it local} density of states at the location of tunneling. The first requirement could be violated if there is a gap or pseudo-gap in the tunneling density of states near zero energy, caused by long-range Coulomb interactions. 
The presence or absence of such a gap should depend only on the character of long-range interaction and the screening properties of the 2DESs, not on the microscopic nature of the tunneling process. Thus, for simplicity, we study a model problem in which an electron tunnels from a localized state near the Probe 2DES to a localized state near the Subject 2DES, and where the only coupling to the 2DESs is through the Coulomb interaction. This has all the same Coulomb physics as the more general problem in which the tunneling electron goes directly from one 2DES to the other. However, in the latter case there are potential complications related to the fact that the tunneling electron is not distinguishable from the electrons that are doing the screening; the justification for ignoring exchange effects in this latter problem has been discussed by Levitov and Shytov.\cite{Levitov}

If transport in the two 2DESs is diffusive, the well known ``zero-bias anomaly'' occurs as a consequence of the inefficient screening of charge in 2D. Specifically, in 2D the conductivity has units of velocity and hence the Coulomb energy $E(t) \sim e^2/R(t)$ associated with adding a charge to a 2D system decays with time in proportion to the screening radius $R(t) \sim \sigma t$. The path integral formulation of the problem results in an action
that logarithmically diverges at small bias for tunneling into such a system. The result is a strong suppression of the tunneling rate near zero-bias,\cite{Altshuler,Levitov,Dynes,Imry} violating criterion (i). Moreover, the tunneling rate has a dominant contribution from long-distance physics, violating criterion (ii).

Our central result states that in the clean limit (infinite mean free path $\ell$),
even in 2D, screening is sufficiently efficient to make the tunneling action at zero-bias finite and hence no zero-bias anomaly occurs. In this regime,
the tunneling rate can be calculated perturbatively and is proportional to the local density of states. Our results indicate that using VSTM to probe the low energy local density of states should be feasible, if the 2DES of interest is clean enough.

Naturally, in any real system, $\ell$ is never infinite.  The screening at asymptotically long distances, and hence the tunneling spectrum at asymptotically low energies, is always diffusive.  Therefore, at low enough energies,   the pseudo-gap behavior of Levitov and Shytov will be recovered.\cite{Levitov}  The crossover between ballistic and diffusive screening occurs on length scales, $\ell$, and hence affects the  tunneling spectrum at energies below
 $E_{co}\sim e^{2}/(\epsilon \ell) = A E_F (1/r_s) (\sigma_Q/ \sigma)$, where $\epsilon$ is the dielectric constant of the semiconductor, $E_F$ is the Fermi energy, $\sigma$ is the conductivity of the screening electron gas, $\sigma_Q = e^2/h$, and $A=2\sqrt{2\pi}$.

The outline of this paper is as follows: In section II, we present the model, which treats the tunneling electron as a two state system and the remaining electrons in the 2DES that interact with the tunneling electron as the ``bath'' degree of freedom. In section III, we calculate the tunneling rate to lowest order in the tunneling matrix element. Finally, in section IV, we discuss the implications of our results.

\section{The Model}\label{The Model}
\begin{figure}[ht]
\includegraphics[width=0.4\textwidth]{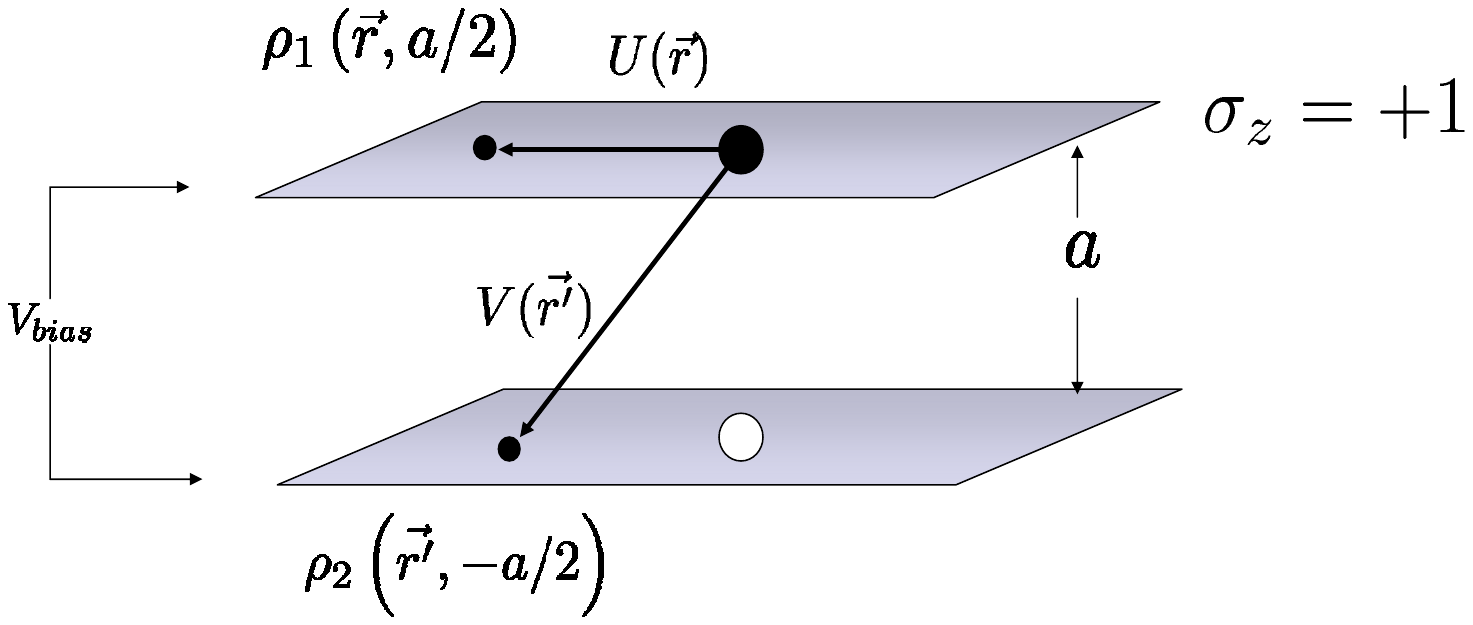} 
\caption{A tunneling electron in state $\sigma_z=+1$ in the upper layer (probe 2DES), interacting with charge density at position $\vec{r}$ in the same layer and charge density at position $\vec{r}'$ in the lower layer (Subject 2DES).}
\label{TunnTheory}
\end{figure}

We consider a simple model that captures essential aspects of the VSTM setup sketched in Fig. \ref{TunnExpt}. Our model consists of two 2DESs characterized by 2D Fermi liquids with electron densities, $\rho_1$ and $\rho_2$, separated by a distance $a$. A voltage bias, $V_{\text{bias}}$, is applied across a single tunneling center at the origin (see Fig. \ref{TunnTheory}). We treat the tunneling electron as a two-state system represented by $\sigma_z=\pm 1$ in the limit of a small bare tunneling matrix element $\Delta$. 
The tunneling electron interacts with the density fluctuations of the 2DESs via a Coulomb interaction.

In the ballistic transport limit, the action for this system is
\begin{equation}
S\left[ \rho _{1},\rho _{2,},\sigma _{z}\right] =S_{\sigma
}+
S_{\sigma ,
\rho}  \label{action}
\end{equation}
Here $S_{\sigma}$ is the bare tunneling action in the absence of any interactions and $S_{\sigma,\rho}$ is the action for the rest of the (``bath'') electrons, which we treat in the context of linear response theory:
\begin{equation}
S_{\sigma }=\frac{1}{2}V_\mathrm{bias}\sigma _{z}-\frac{1}{2}\hbar \Delta \sigma _{x},
\label{Ssig}
\end{equation}
\begin{eqnarray}
S_\mathrm{\sigma, \rho
}&=&\int \frac{d\omega }{\left( 2\pi \right) }\int \frac{d^{2}q} {\left(
2\pi \right) ^{2}}\Big [ {\bf \rho}^\dagger {\bf K} {\bf \rho}+ {\bf \sigma}^\dagger {\bf V} {\bf \rho}\Big ],
\end{eqnarray} 
where  ${\bf \sigma} =(1/2) <[1+\sigma_z(q,i\omega)],[1-\sigma_z(q,i\omega)]>$, ${\bf \rho} = <\rho_1(q,i\omega),\rho_2(q,i\omega)>$, and
\begin{eqnarray}
{\bf K}=&& \!\left( 
\begin{array}{cc}
\chi _{1}^{-1}\left( q,i\omega \right) & V\left( q\right) \\ 
V\left( q\right) & \chi _{2}^{-1}\left( q,i\omega \right) 
\end{array}\right) \\
 {\bf V} =&& \!\left(  \begin{array}{cc}U_{1}\!\left( q\right)\! -\!V\!\left( q\right) & 0 \\ 0 & U_{2}\!\left( q\right)\! -\!V\!\left( q\right) \end{array}\right).
 \nonumber
\end{eqnarray}
 $\chi_i$ denotes the density correlation function in each layer $i$, $V(q)$ denotes inter-layer interaction and ${\bf V}$ is the coupling between the tunneling electron and the ``bath'' electrons through the intra-layer ($U(q)$) and inter-layer ($V(q)$) Coulomb interaction. For simplicity we restrict ourselves to the symmetric case $\chi _{1}^{-1}\left( q,i\omega \right) = \chi _{2}^{-1}\left(q,i\omega \right) = \chi^{-1}(q,i\omega) $ and $U_{1}\left(q\right) = U_{2}\left( q\right) = U(q)$, although the general case can be treated in an identical fashion.

Since Eq. \eqref{action} is quadratic, $\rho_i$'s can be readily integrated out to yield 
\begin{equation}
S\left[\sigma _{z}\right] =\int \frac{d\omega }{2\pi }\left[
\frac{1}{2}V_{\text{bias}}\sigma _{z}\left( i\omega \right) -\frac{1}{2}\Delta \sigma
_{x}\left( i\omega \right) \right] + S_{0},\label{SsigZ}
\end{equation}
where
\begin{equation}
S_{0}\left[\sigma _{z}\right] \equiv -\int \frac{d\omega d^2 q}{(2\pi)^3 }\frac{\left[ (U\left( q\right)-V\left( q\right) \right] ^{2}\left\vert \sigma _{z}\left( i\omega \right)
\right\vert ^{2}}{ \chi^{-1}\left( q,i\omega \right) +V\left( q\right) }.  \label{Sowq}
\end{equation}
Here, the effect of correlations in the 2DESs is encoded in $\chi(q,i\omega)$. We treat the correlation effects at zero temperature, through RPA, in the rest of this paper; however, the form of the action in Eqs. (\ref{SsigZ}-\ref{Sowq}) is more general. The finite temperature calculation is considerably messier but not substantively different. The temperature enters the problem primarily through the modification of the screening properties of the conducting planes. This means that at temperatures low compared to the Fermi energy, where the experiments of interest will be carried out, the effects of finite temperature are expected to be not only continuous, as in the polaron problem,\cite{Silbey} but small as well. The semiclassical results of Levitov and Shytov\cite{Levitov} can be reproduced in this formalism if $\chi$ is taken to be the susceptibility of a diffusive 2DES instead of a ballistic 2DES.

\section{Perturbative Calculation of the Tunneling Rate}\label{Perturbative Calculation}
\subsection{Evaluation of the Action}
We evaluate the action Eqs. (\ref{SsigZ}-\ref{Sowq}) using the RPA expression for $\chi$ in the clean limit:
\begin{equation}
\chi\left( q,i\omega \right)\approx\frac{\chi _{2D}^{o}(q,i\omega )}{1-U\left( q\right)\chi _{2D}^{o}(q,i\omega )},
\label{RPA}
\end{equation}
where the bare density correlator at zero temperature is
\begin{equation}
\chi _{2D}^{o}(q,i\omega )=-\nu _{o}\left( 1-\frac{|\omega |}{\sqrt{\omega
^{2}+(v_{F}q)^{2}}}\right).
\end{equation}
$\nu _{o}=k_{F} / \pi v_{F}$ is the bare density of states per volume at the Fermi surface. Furthermore, we assume that the distance $a$ between the two 2DESs sets the shortest length scale and hence serves as the UV momentum cutoff. 
\footnote{In addition to the distance between the two 2DESs, $a$, other length scales which could affect the short-distance physics include $k_F^{-1}$ and the transverse thickness of the 2DESs.  We assume that these other length scales are at least not much longer than $a$ so that, since we are focusing on long-distance physics, we can treat $a$ as the only short-distance cutoff.}
We then make use of the approximate expression $U\left( q\right) - V\left( q\right) \approx 2\pi ae^{2}+O\left(\left(\left\vert q\right\vert a\right)^{2}\right)$ to lowest order in $|q|a$. This makes it possible to explicitly perform the $q$ integral in Eq. \eqref{Sowq} to yield 
\begin{widetext}
\begin{equation}
S_{\rm eff}(\tau)=S_{\sigma}+\frac{1}{2}\int_{1/\tau }^{1/\tau _{0}}\frac{d\omega }{
2\pi }\left\vert \sigma
_{z}\left( i\omega \right) \right\vert ^{2} \kappa(\omega),
\ \ \  \ \kappa(\omega)\equiv \frac{\alpha^2\omega^2}{4\pi^2 \nu _{o}v_{F}^{2}(1+\alpha )}
\left\{ \left(\frac{v_F}{a\omega}\right) ^{2}
-\frac{2\alpha \log \left[ \left(\frac{v_F}{a\omega}\right)\left( 1+\alpha \right)^2 \right]}{\left( 1+\alpha \right) } \right\}, 
\end{equation}
\label{Seff}
\end{widetext}
We define the dimensionless parameter $\alpha\equiv 2\pi ae^{2}\nu _{o} =2\pi a/a_B$ and the UV frequency cutoff $1/\tau_0 = v_F/a$. Applying the analysis of Levitov and Shytov\cite{Levitov} to this effective action, it is easy to see that the accommodation time and the action at zero-bias are finite; therefore, the tunnelling rate can be computed perturbatively in $\Delta$ as we do explicitly in the next section.

\subsection{Tunneling Rate}
The effective action $S_{\rm eff}$ in section (\ref{Seff}) is of the same general form as for the ``spin-boson'' problem, in which the heat bath is treated as a collection of Harmonic oscillators. 
The heat bath of phonons is typically defined in terms of a spectral distribution function, $J(\omega)$, which is simply the Hilbert transform of the kernel $\kappa(\omega)$ in section (\ref{Seff}):
\begin{equation}
\kappa\left( \omega \right) =\frac{1}{\pi }\int_{0}^{\infty
}d\omega' \left( \frac{\omega' }{\omega ^{2}+\omega^{\prime 2}}%
\right) J\left( \omega' \right),\label{K}
\end{equation}
where
\begin{equation}
J\left( \omega \right) =A\omega^{ 2}e^{-\omega
/\Omega},\label{J}
\end{equation}
and
\begin{equation}
A=\frac{1}{4\pi \nu _{o}v_{F}^{2}}\left( \frac{\alpha }{1+\alpha }\right)
^{3}\hspace{5mm},\Omega=\frac{v_{F}\left( 1+\alpha \right) }{a}.\label{Aom}
\end{equation}
Changes in the form of the high frequency cutoff in the spectral distribution function, $J(\omega)$, result in a frequency independent additive correction to the kernel, $\kappa(\omega)$. In turn, this additive constant produces only an (unimportant) additive correction to the ground-state energy that is not involved in the dynamics.
The low frequency behavior $J\sim \omega^x$ is conventionally classified\cite{Leggett} as ``superOhmic'' for $x >1$ (the present case) where perturbation theory is applicable, "Ohmic" for $x=1$ (which is obtained in the diffusive case), and ``subOhmic'' for $ x < 1$ which requires non-perturbative methods.

With the spectral function in hand, following the steps of Ref. \onlinecite{Leggett}, the tunneling matrix element to second order is 
\begin{align}
\tau ^{-1}\left(V_{\text{bias}}\right)&= \pi^2\hbar\widetilde{\Delta}^2
\delta \left( V_{\text{bias}}\right)\label{tunnrate}\\
&+\!\frac{\pi\hbar 
\widetilde{\Delta}^2}{\varepsilon } \!\sqrt{\!\frac{\varepsilon }{2V_{\text{bias}}}}
I_{1}\left( \!\sqrt{\frac{V_{\text{bias}}}{\varepsilon }}\!\right)\! e^{-V_{\text{bias}}/\hbar \Omega}. \nonumber
\end{align}
Here $I_{1}(x)$ is a modified Bessel function of the first kind, $\varepsilon\equiv\frac{\pi\hbar^{2}}{2A}\propto\varepsilon_{F}$ has the dimension of energy, and $A$ and $\Omega$ are defined in Eq. $\left( \ref{Aom}\right)$. We note that the effect of Coulomb interaction enters the tunneling rate through the renormalized tunneling matrix element
\begin{equation}
\Tilde{\Delta}=\Delta \exp \left[ -\frac{\sqrt{2}r_{s}}{2\pi }\left( 
\frac{1}{\frac{1}{2}\left( \frac{a_{B}}{a}\right) +1}\right) ^{2}\right].
\label{tDelta}
\end{equation}
$r_{s}=\left(1/n\pi\right)^{1/2}a_B^{-1}$ is the ratio of the Coulomb interaction energy to the kinetic energy, $n$ is the electron density, and $a_{B}$ is the effective Bohr radius. Tunneling is suppressed for lower density, i.e. for larger $r_s$. The most notable feature of our results in Eq. \eqref{tunnrate} and Eq. \eqref{tDelta} is the existence of an ``elastic'' term proportional to the Dirac delta function $\delta\left( V_{\text{bias}}\right)$, 
which dominates the tunneling rate in the $V_{\text{bias}}\rightarrow0$ limit. 
This term is absent in the Ohmic and sub-Ohmic cases due to the vanishing overlap (infrared catastrophe) between the $\sigma_z=\pm1$ unperturbed ground states. This proves the existence of a finite tunneling amplitude at zero bias. To illustrate this point, consider tunneling at T=0 not between two individual states but rather between two systems with
density of states $\rho_1(E)$ and $\rho_2(E)$, respectively, to model local tunneling between two Fermi liquids. The tunneling current for $V>E_F$ is then given by
\begin{eqnarray}
I(V) \sim && \int_{0}^{V-E_F} dE_1 \rho_1(E_F+E_1) \nonumber \\
&&\times\int_0^{E_1}dE_2\rho_2(E_F+E_2)\tau^{-1}(E_1-E_2).
\end{eqnarray}
If we assume particle hole symmetry, $\rho_j(E_F+E_j)=\rho_j(E_F-E_j)$ for $j=1,2$, we expect the tunneling current to be odd with respect to $V-E_F$. In addition, if we assume tunneling into a constant density of states $\rho_1(E_F+E_1)=\rho_1$ and $\rho_2(E_F+E_2)=\rho_2$ and combine the tunneling results for $V>E_F$ and $V<E_F$, then the tunneling current is
\begin{eqnarray}
I(V)\sim&& \rho_1\rho_2\pi^2\hbar\tilde\Delta^2 (V-E_F)[1 +\frac {1}{\sqrt{32\pi^2}}\left(\frac{|V-E_F|}{\varepsilon}\right) \nonumber \\
&&+ {\cal O}\left( \frac{|V-E_F|^2}{\hbar\Omega\varepsilon}\right)].
\end{eqnarray}

\section{Summary}
In general, the zero-bias anomaly in tunneling into 2DESs reflects the qualitative effects of Coulomb interactions on the tunneling process. While these effects are interesting in their own right, in the context of VSTM they could represent a barrier to obtaining information on single-particle density of states. Through an explicit
calculation we have shown that in a system where in-plane transport is ballistic and screening is efficient, tunneling is only modestly suppressed by Coulomb effects even in the limit of zero bias. This implies that VSTM will be capable of probing the low energy physics of clean 2DESs through tunneling. The main purpose of the current paper was a proof of principle, hence we limited ourselves to the simplest possible application of VSTM. There are many other systems to which VSTM might be applied where other considerations may be necessary, including tunneling in a magnetic field and tunneling into a non-Fermi liquid. These issues will be the subjects of future studies.

The authors would like to thank Adam Sciambi, Matt Pelliccione, and Mike Lilly for valuable discussions. The Virtual STM concept was developed with support from the Center for Probing the Nanoscale, an NSF NSEC, grant PHY-0425897. The present work was supported by the Department of Energy, through SIMES at the SLAC National Accelerator Laboratory, under DE-FG02-06ER46287 and DE-AC02-76SF00515. K. Luna acknowledges support through a Lucent Bell Labs Graduate Fellowship.

\bibliographystyle{plain}
\bibliography{VSTMbib}

\end{document}